\documentclass[11pt]{article}

\usepackage{moriond,epsfig}

\def\Journal#1#2#3#4{{#1} {\bf #2}, #3 (#4)}

\def\PLB{{\em Phys. Lett.}  B}

\def\PRD{{\em Phys. Rev.} D}

\def\be{\begin{equation}}

\def\ee{\end{equation}}

\def\bea{\begin{eqnarray}}

\def\eea{\end{eqnarray}}

\begin{document}

\vspace*{2cm}
\rightline{DO-TH/04-05}
\vspace*{2cm}
\title{DELINEATING THE (UN)POLARIZED PHOTON CONTENT OF THE  NUCLEON}

\author{ A. MUKHERJEE and C. PISANO}

\address{Instit\"ut fur Physik, Universit\"at Dortmund,\\
D 44221 Dortmund, Germany}

\maketitle
\abstracts{We investigate the QED Compton process (both elastic
and inelastic) in unpolarized and longitudinally polarized
electron-proton scattering. The cross section can be expressed in terms of
the equivalent photon distribution of the proton. We provide the necessary
kinematical cuts to extract the photon content of the proton at
HERMES and eRHIC. We point out that such a process can give valuable
information on $g_1(x_B, Q^2)$ in the small $x_B$, broad $Q^2$ region at
eRHIC and especially in the lower $Q^2$, medium $x_B$ region in fixed target
experiments.}

\section{Introduction}


QED Compton process (QEDCS) in the scattering $ep \rightarrow e\gamma X$ has
a distinctive experimental signature: both the outgoing electron and photon
are detected at large polar angles and their transverse momenta almost
balance each other, with little or no hadronic activity at the detectors
\cite{blu,ruju}. QEDCS in unpolarized $ep$ scattering has long been
suggested as an excellent channel to measure the structure function
$F_2(x_B, Q^2)$ and also to extract the unpolarized photon content  of the
proton in the equivalent photon approximation (EPA) \cite{blu,ruju,kessler}.
In fact, this has been recently analyzed by members of the H1 collaboration
at HERA \cite{lend}. Improved kinematical constraints have been
suggested in \cite{pap1,pp2} for a more accurate extraction of the unpolarized
 photon distribution.
The polarized photon content of the nucleon consists of two components, 
elastic and inelastic, like its unpolarized counterpart \cite{gpr1,gpr2}. 
Recently we showed that when the virtuality of 
the exchanged photon is small, the 'exact' 
polarized QEDCS cross section is expressed in terms of the
polarized equivalent photon distribution of the  proton \cite{pap3}. 
We gave the necessary kinematical cuts to extract the polarized photon 
distribution at HERMES and eRHIC by using QED Compton peak;  QEDCS can also provide valuable
information on $g_1(x_B, Q^2)$ in small $Q^2$, medium $x_B$ region at HERMES
and over a broad range of $x_B$, $Q^2$ at eRHIC. Here we report on our main
results.       
\section{QED COMPTON SCATTERING CROSS SECTION AND THE EPA}
We consider the  process shown in Fig. 1. $X$ is a generic hadronic system
with momentum  $P_{X}=\sum_{X_i} P_{X_i}$. 
For elastic scattering $P_X=P'$ and $X$ is a proton.
\begin{center}
\epsfig{figure= 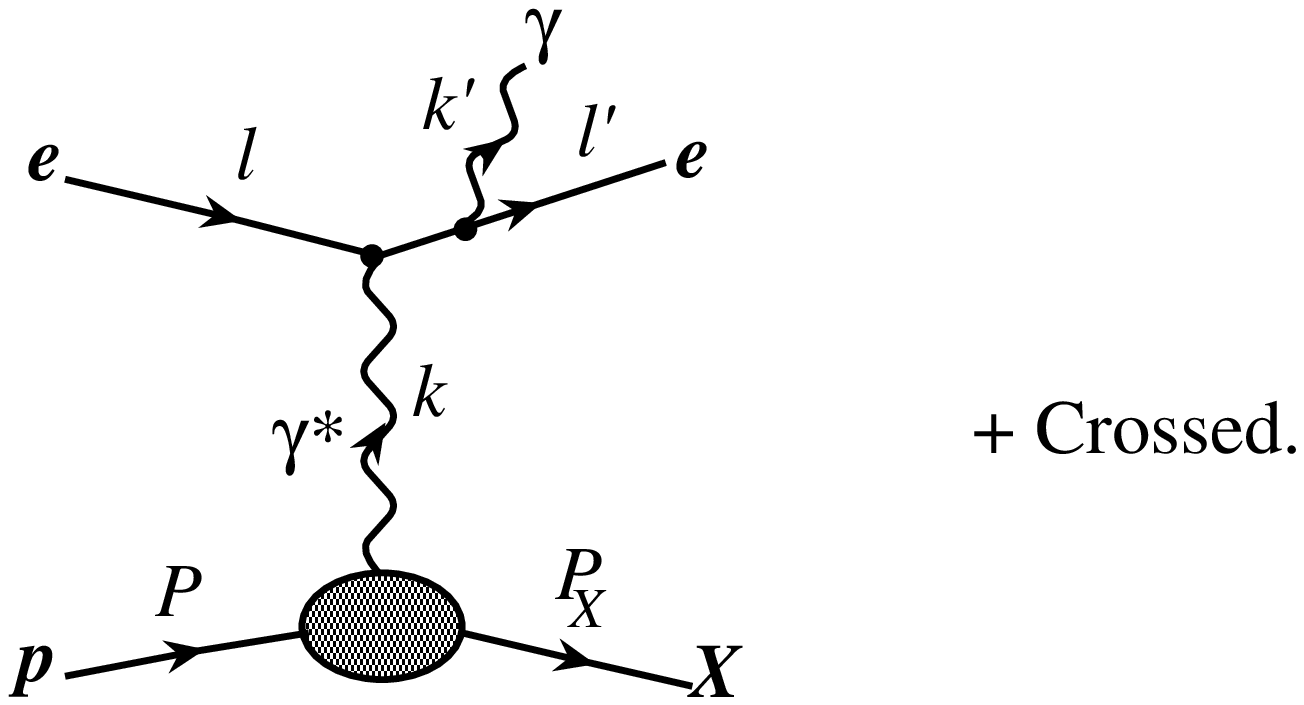, width=14cm, height= 5.5cm}\\
\end{center}  
\begin{center} 
\parbox{10.0cm}   
{{\footnotesize
 Fig. 1:  Feynman diagrams for the QED Compton process (QEDCS). $X \equiv p$
(and $P_X \equiv P'$) corresponds to elastic scattering.}}
\end{center}
\vspace{0.2cm}
We introduce the invariants
\bea
S=(P+l)^2, ~~~~\hat s=(l+k)^2, ~~~~\hat t=(l-l')^2, ~~~~t=k^2,
\label{invar}
\eea
where $k$ is the 4-momentum of the virtual photon.
The photon in the final state is real, $k'^2=0$.  We take the proton mass to
be $m$. The cross section can be calculated in a covariant way \cite{pap1},
both in the elastic and inelastic channels and it can be shown that in the
limit $S \gg m^2$ and $\hat s \gg |t| $, one can approximate the cross section
as
\bea
\sigma (S) \approx \sigma^{\mathrm{EPA}} =
\int_{x_{\mathrm{min}}}^{(1-{m/ 
\sqrt S})^2}\, dx \,\int_{m_e^2 -\hat s}^0
d\hat t  \,\gamma (x, x S) \,{d \hat \sigma (x S, \hat t)
\over d\hat t} ,
\label{epael}
\eea
where $x={\hat s/ S}$ and $\gamma (x, x S)$ is the equivalent 
photon distribution of the proton \cite{blu,ruju,kniehl,gpr1,gpr2}, which
has an elastic and an inelastic component; ${d \hat
\sigma (\hat s, \hat t)\over d\hat t}$ is the real photoproduction cross
section. 

When the incident electron and proton are both longitudinally polarized, the
cross section in the elastic channel  becomes \cite{pap3}
\bea
\Delta \sigma_{\mathrm{el}} &=& {\alpha\over 8 \pi (S-m^2)^2} \int_{m_e^2}^{(
\sqrt{S}-m)^2} d\hat s \int_{t_{\mathrm{min}}}^{t_\mathrm{max}} {dt\over
t} \int_{\hat t_{\mathrm{min}}}^{\hat t_{\mathrm{max}}} d\hat t \int_{0}^{2 \pi} d \phi\,\, X^A_2(\hat s, t, \hat t) \nonumber\\&&~~\times\,\bigg [ \bigg ( 2 \,
{ S-m^2 \over \hat s-t }-1+{2 m^2\over t} {\hat s - t \over
 S-m^2} \bigg ) G_M^2(t)\nonumber\\&&~~~~~~~~~-2\,\bigg ({ S-m^2 \over \hat s-t }
-1+{m^2\over t} {\hat s-t\over S-m^2} \bigg ) {G_M (G_M-G_E)\over 1+\tau}
\bigg ].
\label{elsigg}
\eea   
$G_E$ and $G_M$ are the proton's electric and magnetic form factors and
$\phi$ is the azimuthal angle of the outgoing $e-\gamma$ system 
in the center-of-mass frame. The
limits of integrations follow from kinematics and are the same as in the
unpolarized case \cite{pap1}.  $X_2^A(\hat s, t, \hat t)$ can be obtained
from the leptonic tensor (see \cite{pap3} for the definition). The cross
section in the inelastic channel is \cite{pap3} 
\bea
\Delta \sigma_{\mathrm{inel}}(S) &=& {\alpha\over 4 \pi (S-m^2)^2} \int_{W^2_{\mathrm{min}}}^{W^2_{\mathrm{max}}} dW^2 \int_{m_e^2}^{(\sqrt{S}-m)^2} d \hat s \int_{Q^2_{\mathrm{min}}}^{Q^2_{\mathrm{max}}}
{dQ^2\over Q^2} {1\over (W^2+Q^2-m^2)} \nonumber \\ &&\,\times\,\,\bigg \{ \bigg [-2  {S-m^2\over
\hat s+Q^2}  +{W^2+Q^2-m^2\over Q^2}+{2 m^2 \over Q^2} \bigg ({\hat
s+Q^2\over S-m^2} \bigg ) \bigg ] g_1 (x_B, Q^2) \nonumber\\&&~~~~~+\,\,{4 m^2\over W^2+Q^2-m^2} g_2 (x_B, Q^2) \bigg \}
\tilde{X}_2^A(\hat s, Q^2),  
\eea
here $\tilde{X}_2^A (\hat s, Q^2) = 2\pi\int d \hat t  X_2^A(\hat s,    
Q^2, \hat t) $, $W$ is the invariant mass of the produced hadronic system
and $Q^2=-t$. The limits of the integrations are the same as in the
unpolarized case and can be found in \cite{pap1}. When $S \gg m^2$ and $\hat
s \gg Q^2$, the cross section is approximated  to a form similar to  
(\ref{epael}) with $\gamma(x,x S)$ replaced by $\Delta \gamma(x,x S)$ which
is the polarized equivalent photon distribution of the
 proton and ${d \hat\sigma (\hat s, \hat t)\over d\hat t}$ replaced by 
${d \Delta \hat
\sigma (\hat s, \hat t)\over d\hat t}$, which is the polarized real
photoproduction cross section. $\Delta \gamma(x,x S)$ has both elastic and
inelastic components \cite{pap1,pap3}. The elastic component of $(\Delta)
\gamma$ is  expressed in terms of the form factors for which the well-known
dipole parametrizations can be used \cite{kniehl,gpr1}. The inelastic
component is expressed in terms of the proton structure functions. This
component is scale dependent and is our main concern here. 
\section{NUMERICAL RESULTS}
In this section, we show our numerical estimates of the QEDCS process for
HERMES and eRHIC kinematics respectively. QEDCS events can be selected by
imposing the following constraints on the energies $E_e'$ and $E_\gamma'$ of
the outgoing electron and photon respectively, and on their polar angles
$\theta_e$, $\theta_\gamma $ (these constraints are similar to the ones used
at HERA for unpolarized scattering):
\bea
E_e', E_\gamma' > 4~ \mathrm
{GeV}^2~~~~~~~~~~~~~~~~~~~~~~~~~~~~~~~~~~~~~~~~~\nonumber\\
0.04 \le \theta_e, \theta_\gamma \le 0.2~~ \mathrm{ (HERMES)}; ~~~~
0.06 \le \theta_e, \theta_\gamma \le \pi-0.06 ~~\mathrm{ (eRHIC)}\nonumber\\
\hat s > 1~ \mathrm{ GeV}^2; ~~~\hat s > Q^2~~~~~~~~~~~~~~~~~~~~~~~~~~~~~~~~~~~~~~~~
\label{cuts}
\eea
For HERMES, the incident electron beam energy is $E_e=27.5 $ ~GeV. For
eRHIC, we have taken $E_e=10$ GeV; $E_p=250$ GeV. The constraints on the
energies and the polar angles of the outgoing particles remove the initial
and final state radiative events \cite{blu,kessler} unrelated to QEDCS.   
The last two cuts basically select the preferable kinematical region
where the EPA is expected to  hold.

\vspace{0.3cm}
\begin{center}
\parbox{7cm}{\epsfig{figure=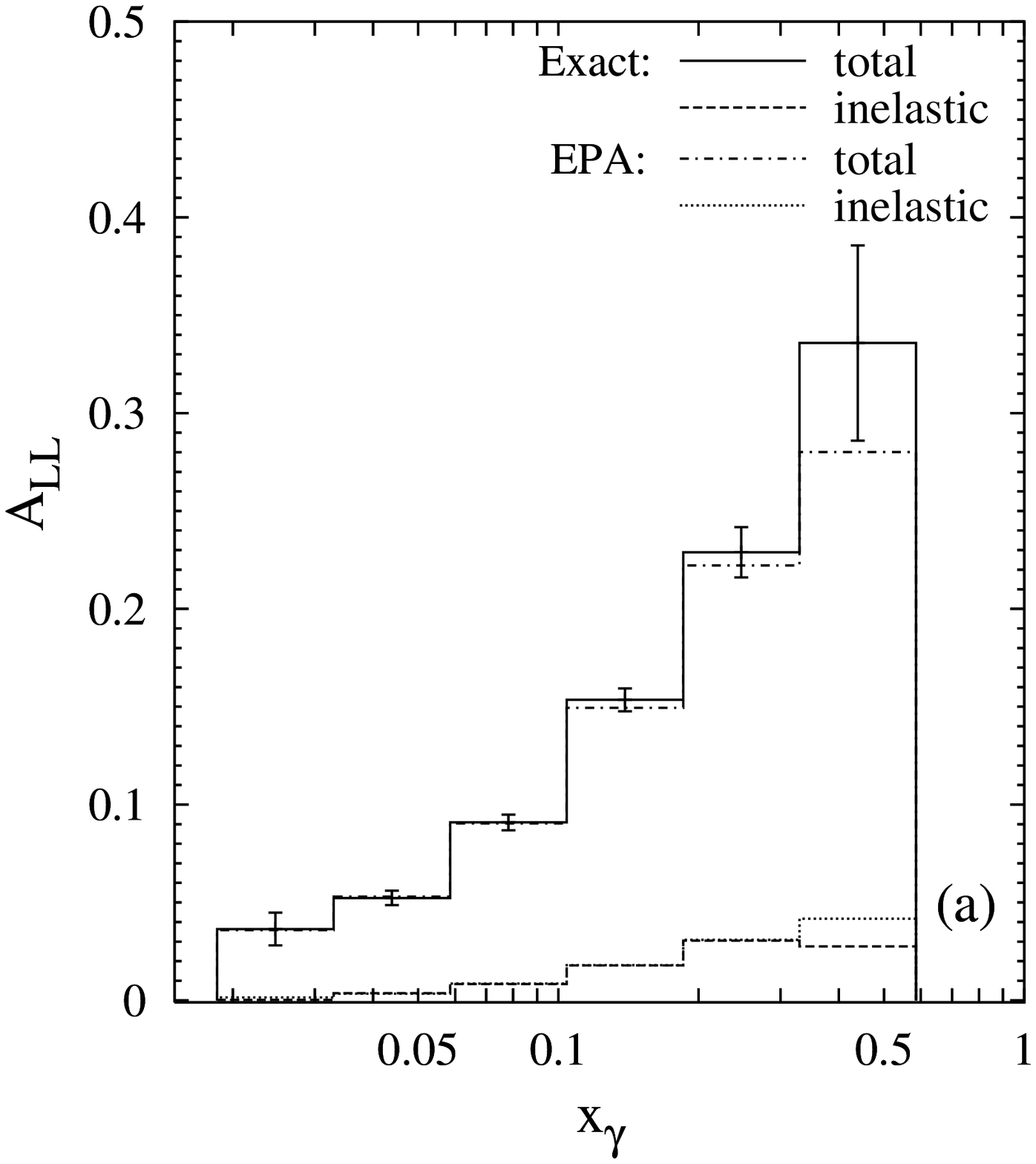,width=7 cm,height=6 cm}}\
\
\parbox{7cm}{\epsfig{figure=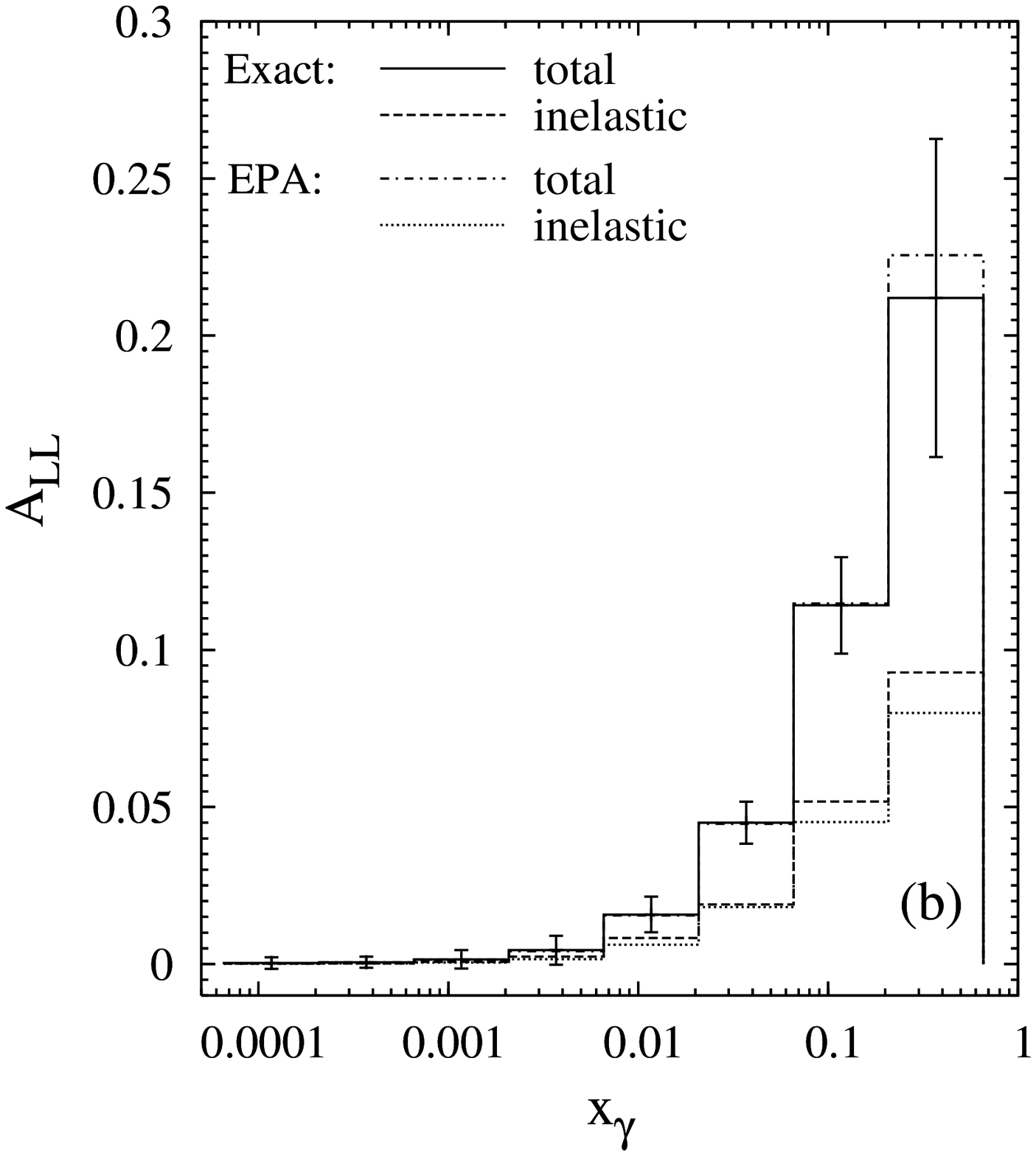,width=7cm,height=6 cm}}\
\
\end{center}
\vspace{0.2cm}
\begin{center} 
\parbox{14.0cm}
{{\footnotesize
Fig. 2 : (a) Asymmetry in bins of $x_\gamma$ at HERMES. (b) Asymmetry at
eRHIC in bins of $x_\gamma$. The
constraints imposed are as in (\ref{cuts}).}}
\end{center}  
\vspace{0.2cm}
The asymmetry $A_{LL}$ is defined as
\bea
A_{LL}={\sigma_{++}-\sigma_{+-}\over \sigma_{++}+\sigma_{+-}}~,
\eea
where the indices $+$ and $-$ refer to the helicities of the incident
electron and proton respectively.
Fig. 2(a) shows the asymmetry for HERMES kinematics in bins of $x_\gamma={l
\cdot k\over P \cdot l}$, which is the fraction of the proton's momentum
carried by the photon. In the EPA, $x_\gamma=x$. The total
(elastic+inelastic) asymmetry shows an excellent agreement with that
calculated in the EPA (shown by the dot-dashed line) in all bins except the
last one for higher $x_\gamma$. The expected statistical error in each bin
is calculated using the formula $\delta A_{LL} \approx {1\over
\mathcal{P}_e \mathcal{P}_p  \sqrt {\mathcal {L}\sigma_{bin}}} $, where
$\mathcal {P}_e$ and $\mathcal {P}_p$ are the polarizations of the
incident lepton and proton respectively, $\mathcal {L}$ is the integrated  
luminosity and $\sigma_{\mathrm{bin}}$ is the unpolarized cross   
section in the corresponding $x_\gamma$ bin. We have taken $\mathcal
{P}_e=\mathcal {P}_p=0.7$ and $\mathcal {L}={1 fb^{-1}}$ for both HERMES and
eRHIC. The asymmetry in the inelastic channel is also shown. The asymmetry
is sizable and can give access to the polarized equivalent photon distribution
 at HERMES. Fig. 2(b)
shows the asymmetry for eRHIC. Here events are observed over a broader range
of  $x_\gamma$, however the asymmetry is very small for small $x_\gamma$
bins, it increases as $x_\gamma$ increases. As in HERMES, good agreement
with the EPA is observed in all but the last bin, where the expected
statistical error is also higher due to the smaller number of events.  
  
The cross section receives a major background contribution coming from
virtual Compton scattering (VCS), when the final state photon is emitted
from the proton side. Particularly important is the inelastic VCS, because 
it affects the determination
of the inelastic component of $(\Delta) \gamma$. The inelastic VCS was estimated in
\cite{pp2,pap3} in an 'effective' parton model (also valid at low $Q^2$).
It was observed that both the polarized and unpolarized VCS contributions
are suppressed in the region $\hat s < \hat S$ where $\hat S= {\hat t (
x_l-x_B)\over x_l}$ with $x_l= {-\hat t \over 2 P \cdot (l-l')}$. Both $\hat
s$ and $\hat S $ are measurable quantities. The interference 
between QEDCS and VCS was found to be suppressed in this region at eRHIC but
not so much at HERMES. However, it changes sign when a positron beam is used
instead of the electron beam, a combination of electron and positron
scattering data can eliminate this contribution.

Finally, we point out that such an experiment can provide valuable
information on the spin structure function $g_1(x_B, Q^2)$ in a kinematical
region not well-covered by fully inclusive experiments (in fact, the
unpolarized structure  function $F_2(x_B, Q^2)$ has been measured by
measuring QED Compton peak at HERA \cite{lend,thesis}) because of its
different kinematics compared to inclusive deep inelastic scattering.
$g_1(x_B, Q^2)$ can be accessed especially at low $Q^2$, medium $x_B$ region
at HERMES and over a very broad $x_B$, $Q^2$ range at eRHIC using the QED
Compton process \cite{pap3}.
\section*{Acknowledgements}
We warmly acknowledge E. Reya and M. Gl\"uck for initiating  this study, as
well as for  many fruitful discussions. 
AM thanks the organizers of the $39$ th Rencontres de Moriond session on QCD and
High Energy Hadronic Interactions for a wonderful and stimulating workshop.
This work has been supported in part
by the 'Bundesministerium f\"ur Bildung und Forschung', Berlin/Bonn.   
 
\end{document}